\documentclass[a4paper,12pt,DIV15]{scrartcl}
\usepackage{amsmath}
\usepackage{amssymb}
\usepackage{authblk}
\usepackage[UKenglish]{babel}
\usepackage{bbm}
\usepackage{bbold}
\usepackage{bm}
\usepackage{dcolumn}
\usepackage{enumerate}
\usepackage{graphicx}
\usepackage[colorlinks=true,linkcolor=red,urlcolor=blue,citecolor=dark-green,menucolor=black,hyperfootnotes,linktocpage]{hyperref}
\usepackage[UKenglish]{isodate}
\usepackage{lscape}
\usepackage{mathrsfs}
\usepackage{mathtools}
\usepackage{multirow}
\usepackage[square,sort&compress,numbers]{natbib}
\usepackage{pifont}
\usepackage{slashed}
\usepackage{subfigure}
\usepackage{units}
\usepackage{xcolor}
\definecolor{dark-green}{rgb}{0.1,0.7,0.3}
\newcommand{\eprint}[1]{\href{http://arxiv.org/abs/#1}{#1}}
\newcommand{\sect}[1]{Sec$.$~#1}
\newcommand{\eq}[1]{Eq$.$~(#1)}
\newcommand{\tab}[1]{Table~#1}
\newcommand{\fig}[1]{Figure~#1}
\newcommand{\nbb}{0 \nu \beta \beta}
\allowdisplaybreaks
\cleanlookdateon
\title{\Large \sffamily\bfseries Lepton Number Violation \\ within  the Conformal Inverse Seesaw}   
\date{} 
\begin{document}
\author{\large Pascal Humbert$^{1}$\footnote{humbert@mpi-hd.mpg.de}  , Manfred Lindner$^{1}$\footnote{lindner@mpi-hd.mpg.de}, Sudhanwa Patra$^{1,2}$ \footnote{sudhanwa@mpi-hd.mpg.de} , Juri Smirnov$^{1}$\footnote{juri.smirnov@mpi-hd.mpg.de} }        
\affil{\footnotesize$^{1}$
Max-Planck-Institut f\"ur Kernphysik, Saupfercheckweg 1, 69117 Heidelberg, Germany \\$^{2}$ Center of Excellence in Theoretical and Mathematical Sciences, \\
Siksha 'O' Anusandhan University, Bhubaneswar-751030, India  } 
\maketitle
\vspace{-1cm}
\begin{abstract}
\noindent
We present a novel framework within the conformal inverse seesaw scheme allowing large lepton number 
violation while the neutrino mass formula is still governed by the low-scale inverse seesaw 
mechanism. This model includes new contributions to rare low-energy lepton number violating processes like neutrinoless double beta decay. We find that the 
lifetime for this rare process due to heavy sterile neutrinos can saturate current experimental limits. 
The characteristic collider signature of the present conformal inverse seesaw scheme includes, same-sign dilepton plus two jets and same-sign dilepton plus four jets. Finally, we comment on the
testability of the model at the Large Hadron Collider since there are new scalars, new fermions and an extra neutral gauge boson with masses around few 100 GeV to few TeV.
\end{abstract}
\section{Introduction}
\noindent
The long standing problem of the 
sensitivity of the Higgs mass to an ultra-violet embedding has motivated many extensions of the Standard Model (SM) at energies close to the Electro-Weak~(EW) scale. However, the first run of the Large Hadron Collider~(LHC) did not find new physics which motivates extensions of the SM without explicit scale. Such extensions would manifest themselves at the colliders in a more indirect  way and would for now look exactly like the SM. 

The hierarchy problem of the Higgs potential can be formulated as an apparent quadratic dependence of observables on an underlying microscopic i$.$e$.$ ultra-violet theory technically addressed as a cut-off dependence. However, it is known that systems close to a critical point and undergoing a phase transition can be described independently of the underlying microphysics \cite{Kadanoff:1971pc}. The essential concept is self-similarity which can arise in form of scale invariance. 

This led to ideas that describe the EW transition in conformal theories where the non-linear realization of the symmetry results in the appearance of scales. It implies that there are no fundamental mass scales a priori in the theory and that consequently all scales we observe emerge dynamically as an effect of quantum interactions. Models of this type have been studied in \cite{Coleman:1973jx, Hempfling:1996ht, Hambye:1995fr, Foot:2007as, Foot:2007ay, Chang:2007ki, Hambye:2007vf, Iso:2009ss, Iso:2009nw, Foot:2010et, Khoze:2013uia, Kawamura:2013kua, Gretsch:2013ooa, Heikinheimo:2013fta, Gabrielli:2013hma, Carone:2013wla, Khoze:2013oga, Englert:2013gz, Farzinnia:2013pga, Abel:2013mya, Foot:2013hna, Hill:2014mqa, Guo:2014bha, AlexanderNunneley:2010nw, Radovcic:2014rea, Khoze:2014xha, Smirnov:2014zga, Salvio:2014soa, Chankowski:2014fva, Okada:2014nea, Guo:2015lxa, Baek:2015mna, Hatanaka:2014tba, Benic:2014aga, Gorsky:2014una, Okada:2014qsa, Okada:2014oda, Khoze:2014woa, Lattanzi:2014mia}. Mathematically, this behaviour is captured by the Coleman-Weinberg mechanism  \cite{Coleman:1973jx}. Since the top quark contribution to the $\beta$-function of the Higgs quartic coupling is large in the SM, 
it turns out that, in order to have Radiative Spontaneous Symmetry Breaking, the particle content of the SM has to be enlarged such that bosonic degrees of freedom become dominant. We will refer to the spectrum of newly added  particles as the Hidden Sector.

In this context it is interesting to discuss SM symmetries in these models. The particle content of the SM is such that lepton and baryon numbers separately and their linear combination $B - L$ are global symmetries. But following fundamental arguments nature should not possess unbroken global symmetries \cite{Kallosh:1995hi}. This would imply that the $B-L$ symmetry is either explicitly broken at a higher scale or that it is gauged. 

The most popular way of explicit lepton number violation (LNV) is to introduce SM singlet fermions $\nu_R$ with lepton number one and a mass term $M_R \bar{\nu}_R \nu_R^c$. As a consequence the $B-L$ symmetry becomes anomaly-free so that it can be gauged, which is another argument in favour of this extension. In the conformal framework, however, explicit mass terms are forbidden. Then the question arises how to properly embed LNV in a conformal model. In a previous work \cite{Humbert:2015epa} we demonstrated how explicit LNV is possible in a conformal model just by interaction terms in the Lagrangian. We found that an extension of the SM gauge group by a $U(1)_X$ local symmetry can lead in this set-up to an inverse seesaw scenario with additional keV-scale Dark Matter. Furthermore, the LNV processes are strongly suppressed in all low-energy observables. In particular there is no neutrinoless double beta decay~($\nbb$) contribution of the new physics. 

In this work we follow an alternative approach and show that it is possible  to break the lepton number spontaneously in the conformal inverse seesaw (CISS). In this scenario the additional local symmetry $U(1)_X$ is identified with $U(1)_{B-L}$. Since the new gauge symmetry does not only operate on the Hidden Sector, but also on SM particles we need to cancel the anomaly contributions from the SM particles as well. This fixes in the case of $B-L$ symmetry the number of SM singlet fermions with $B-L = -1$ to three. The additional fermions have to be organised in pairs vector-like under $B-L$.   

We demonstrate in this work that LNV processes are not suppressed, unlike in the usual inverse seesaw. At the same time the Dark Matter~(DM) phenomenology and the neutrino mass mechanism with sizable active-sterile mixing are preserved as analysed in \cite{Humbert:2015epa}. There, the lepton number was explicitly broken but the $\nbb$ signal was systematically cancelled by pseudo-Dirac contributions. In the present set-up, however, $\nbb$ occurs, even though lepton number is not explicitly broken. It is an impressive fact that the non-linear realization of conformal symmetry forces us to introduce a model with a scalar condensate which is close to the TeV scale, as it is dynamically linked to the EW scale. Therefore, the new physical degrees of freedom are expected to be accessible at the LHC. 

The paper is organised in the following way. In \sect{\ref{sec:Framework}} we discuss how to attain large LNV in the conformal inverse seesaw model. In \sect{\ref{sec:LNV}} we analyse the lepton number violating processes expected in this model and the possibility to distinguish the presented scenario from the CISS with suppressed LNV at the LHC. We summarize our results and conclude in \sect{\ref{sec:Conclusion}}.  
\section{Framework for large lepton number violation \newline  in the conformal inverse seesaw \label{sec:Framework}} 
\noindent
It is clear that the SM needs to be extended in order to explain massive neutrinos. From a theoretical point of view the most obvious way to do so is to introduce new species of neutrinos that can account for mass terms in the Lagrangian of the theory. At the same time neutrino masses have to be tiny compared to the other fermion masses. A popular method to address this issue is the so-called seesaw mechanism that has extensively been studied in various modifications. The canonical (or type-I) seesaw \cite{Minkowski:1977sc, Mohapatra:1979ia} leads to neutrino masses suppressed by a heavy mass scale of the order of $10^{10}~\text{GeV}$ or above, which perfectly can be embedded in e$.$g$.$ a grand unified theory~(GUT). However, such a high mass scale is far beyond reach of particle colliders --- be it existing colliders (LHC) or future colliders. This has led scientists to search for possibilities of a low-scale seesaw mechanism. One possibility to realize this is the inverse seesaw mechanism~\cite{Mohapatra:1986bd,Deppisch:2004fa}. It is characterized by a low lepton number violating mass scale $\mu$ and a heavy mass scale $M$ that can be of order $1~\text{TeV}$, well within reach of the LHC. The inverse seesaw mechanism leads to neutrino masses $\sim (m_D / M)^2 \mu$, where $m_D$ denotes a Dirac mass proportional to the Electro-Weak scale. After this short motivation we can turn our attention to the realization of the aspects just alluded to.
\subsection{The model}
\noindent
The model discussed in this work is based on the conformal inverse seesaw~(CISS) described in our previous work~\cite{Lindner:2014oea,Humbert:2015epa}. In extension we augment the model by a Majorana mass term for the right-handed neutrinos $\nu_R$. In a conformal theory only dimensionless coupling constants are allowed. This means that an explicit mass term for fermions is forbidden, or --- put the other way around --- any fermion mass term present in the Lagrangian has to descend from a Yukawa interaction of the fermion with a scalar field. 

To realize the inverse seesaw pattern we add three right-handed neutrinos $\nu_R$ and two different neutrino species $N_L$ and $N_R$ to the SM. Note that the model is built in a way that it is symmetric under the exchange of $N_L$ with $N_R^c$ and that both fields ought to have the same quantum numbers to guarantee anomaly cancellation in their sector. 

It turns out that, in order to obtain exactly the mass terms that we want to keep, we need to extend the SM gauge group by a new symmetry group that is naturally identified with $U(1)_{B-L}$. To understand this we first give the particle content of the model and the quantum numbers of the fields summarized in \tab{\ref{tab:SM}} and then discuss this particular choice of fields and quantum numbers. 
\begin{table}[h!]
\begin{center}
\begin{tabular}{|c|c|c|c|}
	\hline
			& Field	& $ SU(2)_L\times U(1)_Y$	& $U(1)_{B-L}$	\\
	\hline
	\hline
	Fermions	& $Q_L \equiv(u, d)^T_L$			& $(\textbf{2},~ 1/6)$	& $1/3$	\\
			& $u_R$							& $(\textbf{1},~ 2/3)$	& $1/3$	\\
			& $d_R$							& $(\textbf{1},~-1/3)$	& $1/3$	\\
			& $\ell_L \equiv(\nu,~e)^T_L$	& $(\textbf{2},~  -1)$	&  $-1$	\\
			& $e_R$							& $(\textbf{1},~  -2)$	&  $-1$	\\
			& $\nu_R$						& $(\textbf{1},~   0)$	&  $-1$	\\
			& $N_R$							& $(\textbf{1},~   0)$	&   $3$	\\
			& $N_L$							& $(\textbf{1},~   0)$	&   $3$	\\
	\hline
	Scalars	& $H$							& $(\textbf{2},~ 1/2)$	&   $0$	\\
			& $\phi_2$						& $(\textbf{1},~   0)$	&  $-2$	\\  
			& $\phi_4$						& $(\textbf{1},~   0)$	&   $4$	\\  
			& $\phi_6$						& $(\textbf{1},~   0)$	&   $6$	\\  
	\hline
	\hline
\end{tabular}
\caption{The particle content of the CISS with large lepton number violation. The third and forth columns show the representation of the fields under the Electro-Weak gauge group and, respectively, the quantum number under the new gauge group $U(1)_{B-L}$.\label{tab:SM}}
\end{center}
\end{table}
\newpage

From the particles listed in \tab{\ref{tab:SM}} we obtain the following invariant Lagrangian
\begin{align}
\label{eq:TheModel}
	\mathcal{L}_\text{CISS} 
	&=     i \, \bar{\nu}_R \left( \slashed{\partial} + i\,g_\text{BL} \,Z_\mu^\prime \gamma^\mu \right)\,\nu_R 
	     + i \, \bar{N}_L \left( \slashed{\partial} - 3i\,g_\text{BL} \,Z_\mu^\prime \gamma^\mu \right)\,N_L	
	     + i \, \bar{N}_R \left( \slashed{\partial} - 3i\,g_\text{BL} \,Z_\mu^\prime \gamma^\mu \right)\,N_R 
	\nonumber \\
	&~~~ - \frac{y}{2} \left( \bar{N}_L\,\nu_R \,\phi_4 + h.c.\right) 
	     - \frac{y}{2} \left( \bar{N}_R^c\,\nu_R \,\phi_2 + h.c.\right)
	     - \frac{y_D}{2} \left( \bar{L}\,\tilde{H} \nu_R +h.c. \right) 
	\nonumber \\ 
	&~~~ - \frac{y'}{2} \left( \bar{N}_L\,N_L^c \,\phi_6 +h.c. \right) 
	     - \frac{y'}{2} \left( \bar{N}_R\,N_R^c \,\phi_6 +h.c. \right)
	     - \frac{y_R}{2} \left( \bar{\nu}_R\,\nu_R^c \,\phi_2 +h.c. \right) 
	\nonumber \\
	&~~~ + |\left( \partial_\mu +2 \,i\,g_\text{BL}\,Z'_\mu \right) \phi_2|^2
		 + |\left( \partial_\mu -4 \,i\,g_\text{BL}\,Z'_\mu \right) \phi_4|^2 
	     + |\left( \partial_\mu - 6\,i\,g_\text{BL}\,Z'_\mu \right) \phi_6|^2 
	\nonumber \\
	&~~~ - \frac{1}{4} F_{Z^\prime}^{\mu \nu}F^{Z^\prime}_{\mu \nu} 
	     + \frac{\kappa}{4} F_{Z^\prime}^{\mu \nu}\,F_{\mu \nu}
	     - V \left( H, \phi_2,\phi_4 ,\phi_6 \right)
	     + \mathcal{L}_\text{SM} 
	  \, ,
\end{align}
where $Z_\mu^{\prime}$ denotes the new gauge boson associated with $U(1)_{B-L}$ and $F_{\mu \nu}^{Z^\prime} = \partial_\mu Z_\nu^\prime - \partial_\nu Z_\mu^\prime$ is its field strength tensor. Since $N_L$ and $N_R$ have to have the same quantum numbers to guarantee anomaly cancellation the bilinear combination of both fields ($\overline{N}_L N_R$, $\overline{N}_R N_L$) is a singlet under the model's gauge group. Hence, to avoid such a mass term, we cannot admit a complete singlet scalar field. On the other hand a crucial point of the model is to have a mass term $\overline{\nu}_R \nu_R^c$ for the right-handed neutrinos. Then, in the absence of a scalar singlet, as a consequence it is required that $\nu_R$ is charged under some symmetry group. Finally the Yukawa coupling of the right-handed neutrinos to SM particles $\overline{L} \tilde{H} \nu_R$ forces us to choose a symmetry group that can be reconciled with the SM. The most natural choice for this is to identify the new gauge group with $U(1)_{B - L}$. Note that this identification has led us to introduce the same number of right-handed neutrinos as are present in the SM in order to cancel $U(1)_{B - L}$ anomalies. 
\subsection{Symmetry breaking}
\noindent
The Radiative Spontaneous Symmetry Breaking is similar to the case of the CISS discussed in~\cite{Humbert:2015epa} and leads to the hierarchical vacuum expectation value (vev) structure $\langle \phi_4 \rangle \approx \langle \phi_2 \rangle > \langle H \rangle > \langle \phi_6 \rangle$. The potential yields two vevs of the same order of magnitude, since we assume an exchange symmetry $N_L \leftrightarrow N_R^c$ and $\phi_2 \leftrightarrow \phi_4$, as discussed in our previous paper.  This symmetry leads to equal quartic couplings for the two scalars $\phi_2$ and $\phi_4$. 

The symmetry breaking by the vevs of the scalars naturally gives to masses for fermions and gauge bosons. The neutrino mass terms of the Lagrangian given in \eq{\ref{eq:TheModel}} can be summarized in the Majorana basis as
\begin{equation}
	{\cal L}_{mass}
	= 
	- \frac{1}{2} \bar{\nu^c} \mathcal{M} \nu + \text{h.c.}
\end{equation}
with the neutral lepton mass matrix and flavour basis given by
\begin{align}
	\mathcal{M}
	=
	\begin{pmatrix}
		0						& y_D \langle H \rangle			& 0									& 0									\\
		y_D \langle H \rangle	& y_R \langle \phi_2 \rangle		& y \langle \phi_2 \rangle			& y \langle \phi_4 \rangle			\\
		0						& y \langle \phi_2 \rangle		& y^\prime \langle \phi_6 \rangle	& 0									\\
		0						& y \langle \phi_4 \rangle		& 0									& y^\prime \langle \phi_6 \rangle
	\end{pmatrix} 
	=
	\begin{pmatrix}
		0		& m_D		& 0			& 0		\\
		m_D		& M_R		& M			& M		\\
		0		& M			& \mu_1		& 0		\\
		0		& M			& 0			& \mu_2
	\end{pmatrix} 
    \, ,
	&&
	\nu
	=
	\begin{pmatrix}
		\nu_L	\\
		\nu_R^c	\\
		N_L		\\
		N_R^c
	\end{pmatrix}
	\, .
	\label{massMatrix}
\end{align}
Remember that we assume that $M_R$ is the largest of all elements. Note that in this set-up both $M_R$ and $M$ are proportional to $\langle \phi_2 \rangle \approx \langle \phi_4 \rangle$. This means a hierarchy $M_R \gg M$ between the mass terms must follow from a hierarchy in the Yukawa couplings, $y_R \gg y$.

After spontaneous symmetry breaking, the vevs of the non-SM scalars give a mass to the extra neutral gauge boson $Z^\prime$. The SM Higgs vev generates mass terms for the other neutral gauge bosons $B$ and $W^{3}$. Neglecting the kinetic mixing among the $U(1)$ gauge bosons the neutral gauge boson mass matrix  in the basis $(W_{\mu}^{3},~B_{\mu},~Z_\mu^\prime)$ reads
\begin{eqnarray}
	\mathcal{M}_\text{neutral}^2
	=
	\left(\begin{array}{ccc}
		\frac{1}{4} g^2 \langle H \rangle^2	& -⁠\frac{1}{4} g g^\prime \langle H \rangle^2			& 0	\\
		-\frac{1}{4} g g^\prime \langle H \rangle^2	& \frac{1}{4} g^{\prime^2} \langle H \rangle^2	& 0	\\
		0	& 0	& 8 g^2_{BL} \left( \langle \phi_2 \rangle^2+ 4 \langle \phi_4 \rangle^2 + 9 \langle \phi_6 \rangle^2\right)									\\
\end{array} \right)
\, .
\end{eqnarray}
For the allowed vev hierarchy the physical mass of the $Z^\prime$ is given by
\begin{equation}
	M^2_{Z^\prime}
	=
	8 g^2_{BL} \left(\langle \phi_2 \rangle^2+ 4 \langle \phi_4 \rangle^2 +9 \langle \phi_6 \rangle^2\right)
	\, .
\end{equation}
Additionally, we obtain $M_Z=\frac{1}{2} \langle H \rangle \sqrt{g^2+g^{\prime^2}}$ for the $Z$ boson mass and a massless photon $A$, as in the SM.  
\subsection{Masses and Mixing}
\noindent
In the following we will qualitatively discuss the mass eigenvalues and mixing obtained from the matrix structure shown in \eq{\ref{massMatrix}}. We define the (complete) mixing matrix $U$ via 
\begin{equation}
	\mathcal{M}_\text{diag} 
	= 
	U^\dag \mathcal{M} U^*
	=
	V^\dag \cdot \{ W^\dag \mathcal{M} W^* \} \cdot V^*
	\, ,
	\label{massMatrixDiagonal}
\end{equation}
where $\mathcal{M}_\text{diag} = \text{diag}(m_1,~m_2,~m_3,\ldots)$ contains the physical neutrino masses. As indicated in \eq{\ref{massMatrixDiagonal}} the diagonalization of $\mathcal{M}$ can be carried out in two steps, first a block diagonalization $W$ and second the diagonalization $V$ of the blocks obtained that way. Then we can bring $\mathcal{M}$ into block-diagonal form by transforming
\begin{equation}
	\mathcal{M}_\text{block}
	=
	W^\dag \mathcal{M} W^*
	=
	\text{diag} \left( m_\nu,  m_\text{keV},  m_\text{int}, m_\text{heavy} \right)
	\, . 	
\end{equation}
The matrices $m_\nu,~m_\text{keV},~m_\text{int}$ and $m_\text{heavy}$ denote the active neutrino mass matrix and a keV scale, intermediate scale and heavy scale mass matrix, respectively. Assuming the hierarchy $M_R > M > m_D > \mu_+$, where we have defined $\mu_+ = \mu_1 + \mu_2$, they are proportional to 
\begin{align}
	&m_\nu 
	\sim
	\left( \frac{m_D}{M} \right)^2 \mu_+
	\, ; \quad \mbox{inverse seesaw formula for light neutrinos}
	\, , 
	\nonumber \\
	&m_\text{keV} 
	\sim 
	\mu_+
	\, ,	~~~
	m_\text{int}
	\sim 
	\frac{M^2}{M_R}
	\, ,~~~ 
	m_\text{heavy}
	\sim 
	M_R
\end{align}
up to negligible corrections from the block-diagonalization. We find that the minimal configuration, where this mass pattern is stable, is a model with $(3 + 3 + 2 + 2)$ eigenstates in the flavour basis $\nu$. Examining the spectrum of mass scales in agreement with light neutrino masses we find that for $M_R$ in the few TeV scale we have intermediate scale Majorana neutrinos from a few GeV up to a few 100 GeV. In addition there  is a state in the few keV range  which is a perfect candidate for Dark Matter and could explain the recent observation of a $3.51$ keV X-ray line \cite{Bulbul:2014sua}. 
The DM phenomenology is unaffected by the lepton number violating mass term. Thus the analysis presented in our previous work applies to this model, too. The correct DM relic abundance is assumed to be generated in a freeze-in process. For a more detailed discussion we refer to \cite{Merle:2015oja}. We point out that a slight modification of the model leads to a stable weakly interacting massive DM candidate. If there is no $\phi_4$ scalar with the $B-L$ quantum number $4$ in the theory, $B-L$ breaks to a remnant $Z_2$ symmetry. This symmetry is the reason why one of the fermions with the mass of the order of the $\phi_6$ vev remains stable. This particle is produced in s-channel interactions with the $Z^\prime$ gauge boson and can account for the correct DM relic abundance after a freeze-in. This change does not affect the phenomenology of the LNV and we will thus postpone a detailed analysis.

To discuss the mixing pattern let us define the following hierarchy parameters 
\begin{align}
	\sigma 
	= 
	\frac{M}{M_R} 
	\, ,
	&& 
	\epsilon 
	= 
	\frac{m_D}{M} 
	\, ,
	&& 
	\eta 
	=
	\frac{\mu_+}{M}
	\, . 
\end{align}
Considering the same mass hierarchy as above $M_R > M > m_D > \mu_+$ the block-diagonalization can be put in the approximate form
\begin{equation}
	\mathcal{O}(W^\dag) 
	=
	\begin{pmatrix}
		1				& \epsilon \eta	& \epsilon				& \epsilon				\\
		0				& 0				& \frac{1}{\sqrt{2}}		& \frac{1}{\sqrt{2}}		\\
		\epsilon			& \sigma			& \frac{1}{\sqrt{2}}		& \frac{1}{\sqrt{2}}		\\
		\epsilon \sigma	& 1				& \sigma					& \sigma
	\end{pmatrix}	
	\, . 
\label{mixingMatrix_orderOfMagnitude}
\end{equation}
Note that the elements of $W$ have the same order of magnitude as the elements of the mixing matrix $U = W \cdot V$, since $\mathcal{O}(V) = \mathbbm{1}$. 

The mixing matrix connects the flavour with the mass basis $n \equiv (\nu_\text{SM},~\nu_\text{keV},~N_\text{int},~N_\text{heavy})^T$ via 
\begin{equation}
	\nu_\alpha 
	= 
	U_{\alpha i}^* n_i 
	=
	\sum_{i = 1}^3				U_{\alpha i}^* \nu_{\text{SM} i} 	+
	\sum_{i \in \text{keV}}		U_{\alpha i}^* \nu_{\text{keV} i} 	+
	\sum_{i \in \text{int}}		U_{\alpha i}^* N_{\text{int} i} 		+
	\sum_{i \in \text{heavy}}	U_{\alpha i}^* N_{\text{heavy} i}
	\, .
\label{eq:nuflav}
\end{equation}
Note that $\nu_\text{SM},~\nu_\text{keV}$ denote relatively light neutrino states while $N_\text{int},~N_\text{heavy}$ are relatively heavy. As a result of this new relation for $\nu_{\alpha}$ given in \eq{\ref{eq:nuflav}}, the charged-current~(CC) interaction Lagrangian in the lepton sector becomes
\begin{align}
	{\cal L}_{\rm CC} 
	&= 
	\frac{g}{\sqrt 2} W^\mu 
	\sum_{\alpha=e, \mu, \tau} \overline{\ell_\alpha} \gamma^\mu P_L \nu_\alpha 
	+ {\rm h.c.} 
	\nonumber \\
	&= \frac{g}{\sqrt 2} W^\mu  
	\sum_{\alpha=e, \mu, \tau} \overline{\ell_\alpha} 
	\gamma^\mu P_L \bigg\{ 
	\sum_{i = 1}^3				U_{\alpha i}^* \nu_{\text{SM} i} 	+
	\sum_{i \in \text{keV}}		U_{\alpha i}^* \nu_{\text{keV} i} 	+
	\sum_{i \in \text{int}}		U_{\alpha i}^* N_{\text{int} i} 		+
	\sum_{i \in \text{heavy}}	U_{\alpha i}^* N_{\text{heavy} i}
	\bigg\}
	\nonumber \\
	&~~~ + {\rm h.c.}\, .
\label{eq:ccLagrangian} 
\end{align}
With the help of the CC interaction Lagrangian we are able to calculate the amplitudes for decays as well as for scattering processes. 
\section{Lepton number violation \label{sec:LNV}}
\noindent
To admit a heavy Majorana mass term for the right-handed neutrinos naturally leads to strong LNV. Such lepton number violating physics will manifest itself in new processes not present in the SM. In the following we discuss the major impact of the LNV obtained in our model specifically on $\nbb$ and same sign dilepton signatures at the LHC.\footnote{We will not discuss lepton flavour violation (LFV), as the relevant constraints presented in our previous analysis \cite{Humbert:2015epa} apply here as well. Next generation LFV experiments will test deeper into the parameters space of the model.}  
\subsection{Neutrinoless double beta decay}
\noindent
One possibility of observing LNV is the neutrinoless double beta decay~($\nbb$). It is the (hypothetical) simultaneous decay of two neutrons of the nucleus of an isotope $(A,~Z)$ into two protons and two electrons without the emission of any neutrinos,
\begin{equation}
	\nbb 
	\, : \, 
	(A, \, Z) \rightarrow (A, \, Z + 2)^{++} + 2 e^- 
	\, . 
\end{equation}
The non-observation of such a decay can be interpreted as a lower limit on the halflife of the isotope under investigation. Physically the halflife can be expressed in terms of a phase-space factor $\mathcal{G}_{(A, \, Z)}^{0 \nu}$, a nuclear matrix element $\mathcal{M}_{(A, \, Z)}^{0 \nu}$ and a dimensionless effective parameter $\eta_{\, \text{eff}}^{0 \nu}$ according to 
\begin{equation}
	(T_{1/2}^{0 \nu})^{-1}_{(A, \, Z)} 
	= 
	\mathcal{G}_{(A, \, Z)}^{0 \nu} |\mathcal{M}_{(A, \, Z)}^{0 \nu} \eta_{\, \text{eff}}^{0 \nu}|^2
	\, .
\label{halflife_formula_general}
\end{equation}
The phase-space factor is responsible for the kinematics of the decay and highly energy dependent. The nuclear matrix element~(NME) takes care of the transition of the nucleus into its daughter. Since it describes a multi-particle process this quantity constitutes the largest source of uncertainties in deriving particle physics constraints from the experimental bounds of the halflife. Finally the effective parameter contains the particle physics of the transition $2 d \rightarrow 2 u + 2 e^-$ inside of the involved nucleons.

\begin{figure}[t!]
\centering
\includegraphics[scale=0.5]{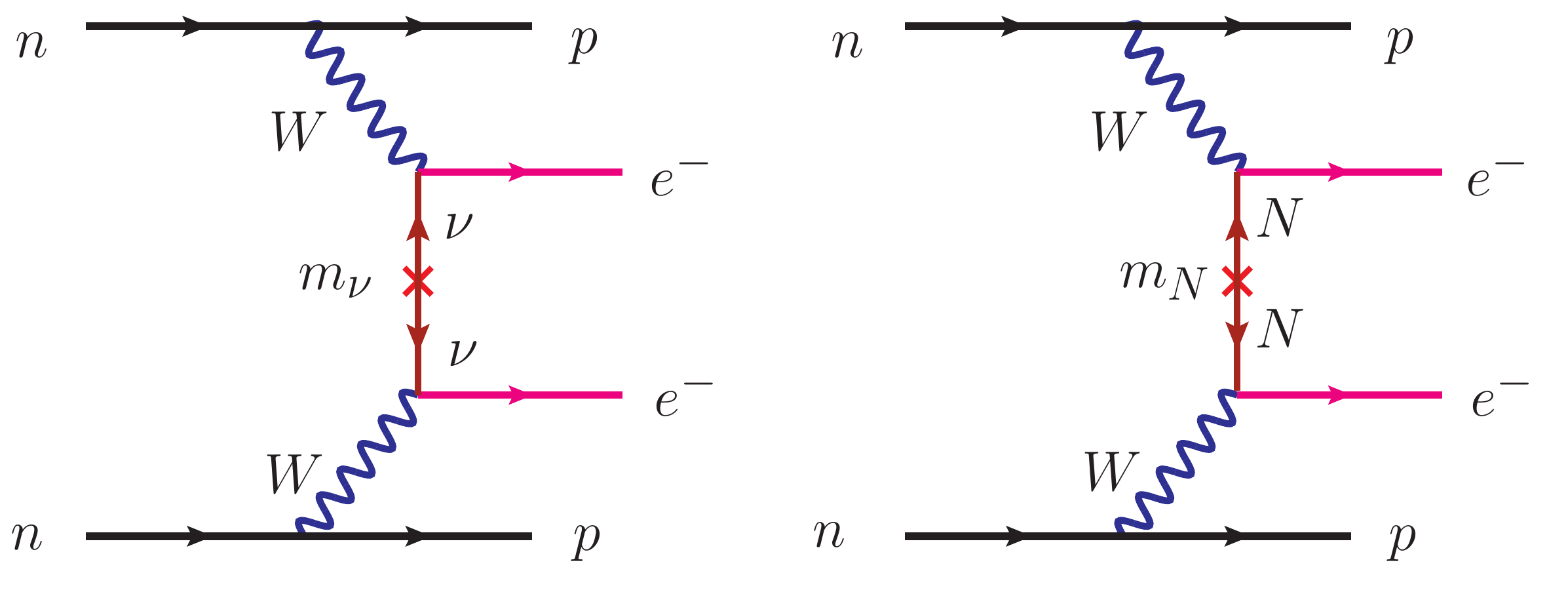}
\caption{Feynman diagrams contributing to neutrinoless double beta decay with $W^- - W^-$ mediation via the exchange of virtual light neutrinos $\nu$ (left panel), and the exchange of virtual heavy neutrinos $N$ (right panel).}
\label{fig:0nubb-ciss}
\end{figure}

From the particle physicist's point of view the observation of $\nbb$ would prove the existence of an (effective) LNV operator. The common explanation --- called the standard mechanism --- is that neutrinos are Majorana particles so that a process as shown in \fig{\ref{fig:0nubb-ciss}} (left panel) is possible. In this case the effective parameter introduced in \eq{\ref{halflife_formula_general}} is given by the $ee$ element of the Majorana mass matrix normalized to the electron mass
\begin{equation}
	\eta_{\, \text{eff}}^{0 \nu} 
	\equiv 
	\frac{m_{ee}}{m_e} 
	= 
	\frac{1}{m_e} \left(	
	\sum_{i = 1}^3 (U_\text{PMNS})_{e i}^2 \, m_i
	\right)
	\, .
\end{equation}
Note, however, that the standard mechanism is not the only way to realize $\nbb$. In principle any new physics that violates lepton number (effectively) by two units can lead to $\nbb$. Additionally it is possible that not only one but several mechanisms give significant contributions to the amplitude of $\nbb$ and lead to interference phenomena.

The process leading to $\nbb$ in our model is the same as the one shown in \fig{\ref{fig:0nubb-ciss}}. The difference to the standard mechanism, however, is that we have additional contributions coming from the new neutrino states. In general one can distinguish between light ($\nu$) and heavy ($N$) neutrino exchange. Let us define the following dimensionless parameters for the exchanges
\begin{align}
	\eta_{\, \nu} 
	&=
	\frac{1}{m_e} \left(
	  \sum_{i = 1}^3 U_{e i}^2 m_i 
	+ \sum_{i \in \text{keV}} U_{e i}^2 m_i
	\right) 
	\approx
	\frac{m_{ee}}{m_e}
	\, ,
	\label{0nubb_light_contribution}
	\\
	\eta_{\, N}
	&=
	- m_p \left( 
	  \sum_{i \in \text{int}} U_{e i}^2 \frac{1}{m_i}
	+ \sum_{i \in \text{heavy}} U_{e i}^2 \frac{1}{m_i}
	\right)
	\equiv
	- \frac{m_p}{m_N}
	\, .
	\label{0nubb_heavy_contribution}
\end{align}
Note that these quantities are normalized to the electron mass, $m_e$, and proton mass, $m_p$, respectively. For the approximation in \eq{\ref{0nubb_light_contribution}} we have taken into account that the mixing of the electron neutrino to the keV states is negligible [$ \mathcal{O}(U_{e \, \text{keV}}^2) \sim 0$, cf$.$ \eq{\ref{mixingMatrix_orderOfMagnitude}}]. The light and heavy neutrino exchange in general have different NME's. We will denote them by $\mathcal{M}_\nu$ and $\mathcal{M}_N$, respectively (see \tab{\ref{tab:nucl-matrix}} for the numerical values).\footnote{
Here and in the following we omit the specification of the isotope $(A, \, Z)$.
}
\begin{table}
\centering
\vspace{10pt}
\begin{tabular}{lcccc}
	\hline 
	\hline 
	Isotope	& $G_{0\nu}[10^{-15} \, {\rm yrs}^{-1}]$	& {$\mathcal{M}_\nu$}	& {$\mathcal{M}_N$}	\\
	\hline
	$^{76}$Ge	& $7.98$		& 3.85--5.82		& 172.2--411.5	\\ 
	$^{136}$Xe	& $59.2$		& 2.19--3.36		& 117.1--172.1	\\ 
\hline \hline
\end{tabular}
\caption{The numerical values of the phase-space factor and nuclear matrix elements taken from \cite{Meroni:2012qf}. Note that the ranges for the nuclear matrix elements correspond to the extremal values given in the reference.}
\label{tab:nucl-matrix}
\end{table} 
The halflife of $\nbb$, \eq{\ref{halflife_formula_general}}, then is given by
\begin{align}
	(T_{1/2}^{0 \nu})^{-1}
	=
	\mathcal{G}^{0 \nu} \left| \mathcal{M}_\nu \, \eta_{\, \nu} 
	+ \mathcal{M}_N \, \eta_{\, N} \right|^2
	\approx
	\mathcal{G}^{0 \nu} \left| \frac{\mathcal{M}_\nu}{m_e} \right|^2 
	\left|m_{ee} - m_e m_p \frac{\mathcal{M}_N}{\mathcal{M}_\nu} m_N^{-1} \right|^2
	\, .
\label{halflife_formula_our_model}
\end{align}
Note that the typical momentum transfer for $\nbb$ is $\langle p^2 \rangle = \left| - m_e m_p \frac{\mathcal{M}_N}{\mathcal{M}_\nu} \right| = (190~\text{MeV})^2$. From the right-hand side of \eq{\ref{halflife_formula_our_model}} we see that we can in general expect interference effects between the light and heavy neutrino contributions. However, in the case where one contribution is dominant compared to the other, the interference between the different mechanisms can be neglected without loss of generality.

\begin{figure}[t!]
\centering
	\includegraphics[width=0.49\columnwidth]{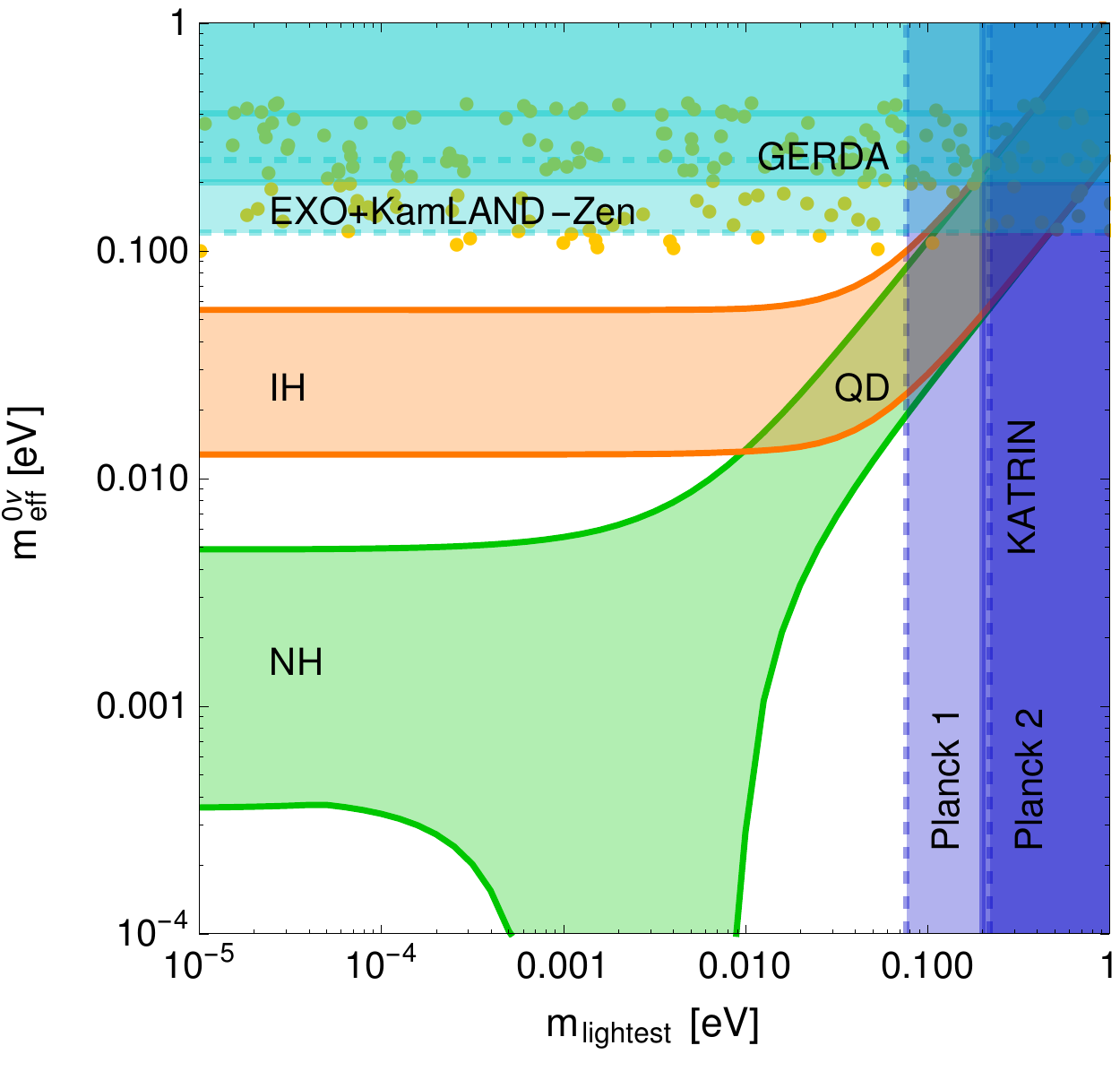}
	\includegraphics[width=0.49\columnwidth]{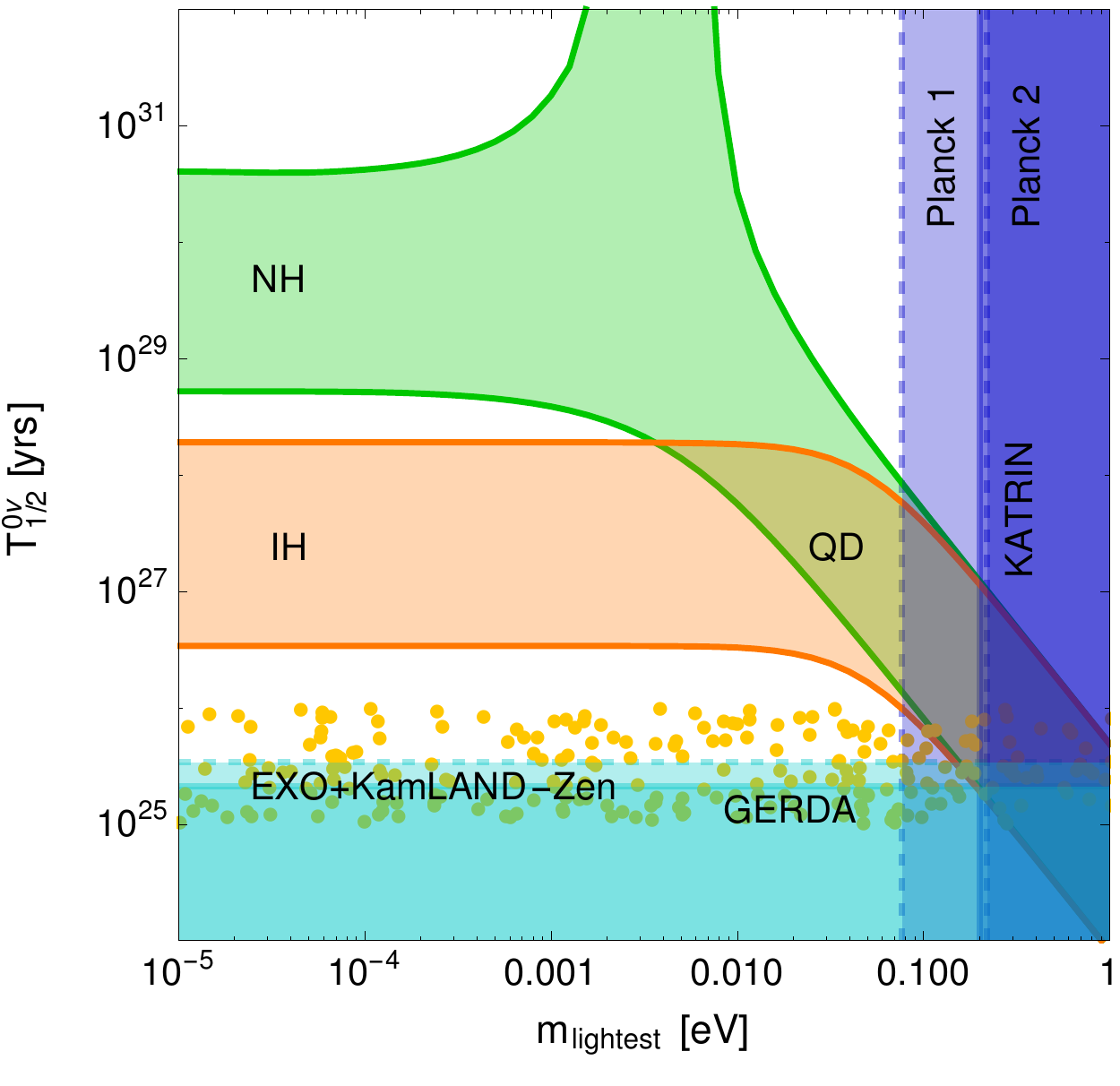}
\caption{Heavy neutrino contributions to the effective Majorana mass (left panel) and the corresponding halflife (right panel) of neutrinoless double beta decay against the lightest neutrino mass displayed as yellow dots. The dots show values that we choose to saturate the experimental limit on the halflife (represented by horizontal lines and the respective shaded areas for the GERDA and EXO+KamLAND-Zen experiments). Note that the seemingly more stringent constraints in the plot of the effective Majorana mass is a result of the uncertainty in the nuclear matrix elements. The limit on the sum of light neutrino masses from cosmological data (Planck 1 and 2) as well as the prospected reach of the KATRIN detector are represented by vertical lines and the respective shaded areas. The green and the red areas, respectively, show the $3 \sigma$ oscillation data allowed ranges in a three-neutrino scheme for normal hierarchy (NH) and for inverted hierarchy (IH), respectively. The quasi-degenerate regime (QD), where NH and IH merge, is indicated.}
\label{plot:hlife-0nubb-ciss}
\end{figure}

In \fig{\ref{plot:hlife-0nubb-ciss}} we plot the effective Majorana mass and the corresponding halflife of $\nbb$ as a function of the mass of the lightest neutrino. The yellow dots are the prediction of a $\nbb$ signal coming from a dominant heavy neutrino contribution in the LNV CISS framework. They show that the current and future experimental limits on the halflife of $\nbb$ can well be saturated, if the LNV heavy contribution is strong enough. In \tab{\ref{tab:0nbb_limits}} we list the current experimental limits on the halflife and the corresponding mass parameter for the isotopes $^{76}$Ge and $^{136}$Xe shown in \fig{\ref{plot:hlife-0nubb-ciss}}.
\begin{table}[h!]
\begin{center}
	\begin{tabular}{c|c|c|c}
	Isotope & $T_{1/2}^{0 \nu}~[10^{25}\text{ yrs}]$ & $m_\text{eff}^{0 \nu}~[\text{eV}]$ & Collaboration \\
	\hline
	$^{76}$Ge	& $> 2.1$	& $< (0.2 - 0.4)$ 	& GERDA~\cite{Agostini:2013mzu} 			\\
	$^{136}$Xe	& $> 1.6$	& $< (0.14 - 0.38)$ 	& EXO~\cite{Auger:2012ar} 				\\
	$^{136}$Xe	& $> 1.9$	& n/a 				& KamLAND-Zen~\cite{Gando:2012zm} 		\\
	$^{136}$Xe	& $> 3.6$	& $< (0.12 - 0.25)$ 	& EXO + KamLAND-Zen combined~\cite{Gando:2012zm} 	
	\end{tabular}
\caption{The current lower limits on the halflife $T_{1/2}^{0 \nu}$ and upper limits on the effective mass parameter $m_\text{eff}^{0 \nu}$ of neutrinoless double beta decay for the isotopes $^{76}$Ge and $^{136}$Xe. The range for the effective mass parameter comes from different calculation methods for the nuclear matrix elements.}
\label{tab:0nbb_limits}
\end{center}
\end{table}
\newpage
\subsection{Probing lepton number violation at colliders}
\begin{figure}[tb]
\centering
\centering
	\includegraphics[width=0.49\columnwidth]{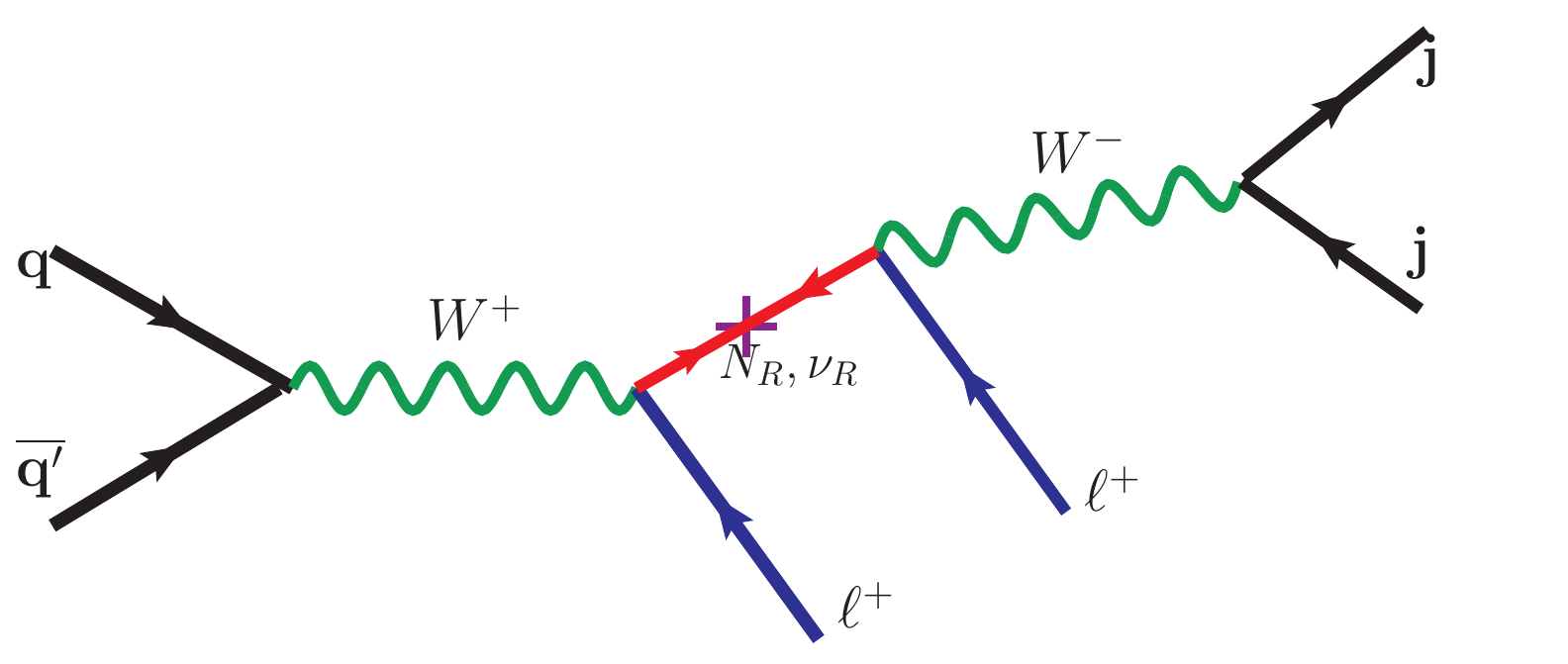}
	\includegraphics[width=0.49\columnwidth]{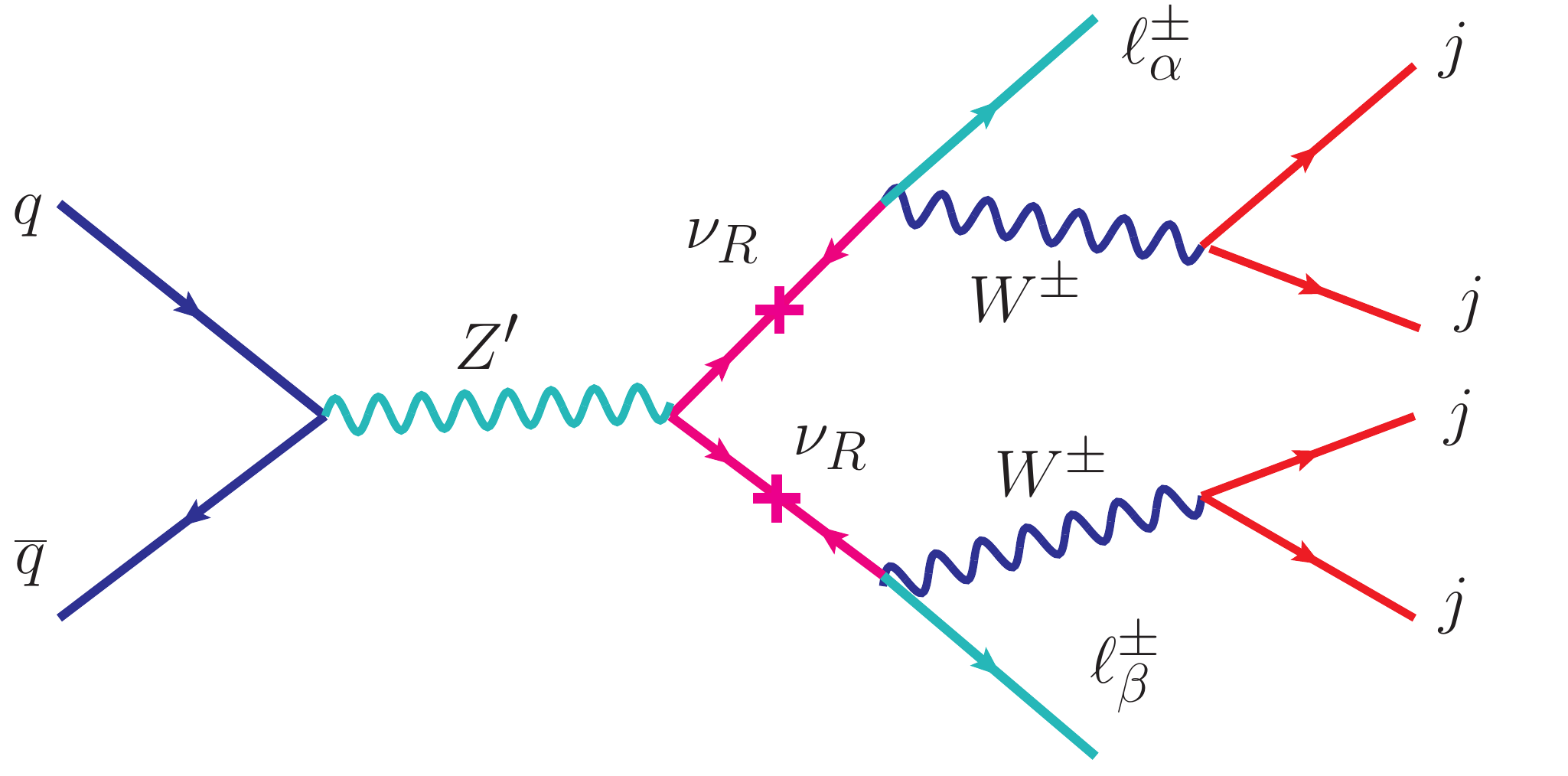}
\caption{Production of heavy neutrinos leading to lepton number violating same-sign dilepton signatures at Colliders. The left panel shows the $pp\to \ell \ell +2j$ process while the right panel displays the process $pp\to Z^\prime \to \ell \ell +4j$. Note that there are additional contributions to the ($\ell^\pm \ell^\pm + 4j$) signal arising from the decay of the $\phi_2$ scalar produced via mixing with the SM Higgs.}
\label{fig:samesign-dilepton}
\end{figure}

\noindent
The characteristic collider signature probing lepton number violation is the same-sign dilepton plus two jets signal ($\ell^\pm \ell^\pm + 2j$) and the same-sign dilepton plus four jets signal ($\ell^\pm \ell^\pm + 4j$), both without missing energy. In the left panel of \fig{\ref{fig:samesign-dilepton}} we illustrate the Feynman diagram for the ($\ell^\pm \ell^\pm + 2j$) signal while the right panel shows the diagram for the ($\ell^\pm \ell^\pm + 4j$) mediated by $Z^\prime$ decay. 

The same-sign dilepton signal is primarily depending upon the large light-heavy neutrino mixing and the mass of the sterile neutrinos, but of course the production mechanism for these processes plays an important role, too. The dependence on the mass is most drastically seen in the different halflife of the heavy neutrino decay. For the ($\ell^\pm \ell^\pm + 2j$) signal there are two distinct cases. If the heavy neutrino mass is larger than $M_W$ the neutrino decays immediately, and most probably into a charged lepton and two jets as shown in the figure. But for masses in the regime of about $5~\text{GeV}$ up to $M_W$ the neutrino will travel some distance before decaying which leads to a displaced vertex of leptons \cite{Izaguirre:2015pga}. Thus, for neutrinos in this mass range we expect the signal to be a prompt charged lepton and a displaced leptonic vertex. Another kinematic observable is the angle between the produced charged leptons. For small neutrino masses $(100~\text{GeV})$ the charged lepton tracks are most likely to be parallel, while for large masses $(800~\text{GeV})$ a back-to-back emission is expected \cite{Almeida:2000pz}. 

The event topology contributing to the ($\ell^\pm \ell^\pm + 4j$) final state is displayed in \fig{\ref{fig:samesign-dilepton}} (right panel). It shows the LNV decay of the $Z^\prime$ boson into two heavy neutrinos with $M_{Z^\prime} = g_\text{BL}\sqrt{\left(8 v^2_2+ 32 v^2_4+72 v^2_6\right)}$. Similar contributions can arise through the decay $pp \to H \to \phi_2 \to \ell^\pm \ell^\pm + 4j$ taking into account the natural mixing of the scalar $\phi_2$ with the SM Higgs boson which is required by the Radiative Spontaneous Symmetry Breaking. Note that since both processes have an s-channel mediator exchange there are two resonances expected in the total invariant mass of these final states corresponding to the $Z^\prime$ and the $\phi_2$ scalar boson. 

Now let us investigate the ($\ell^\pm \ell^\pm + 2j$) signal in the context of the CISS framework with large LNV contribution coming from heavy Majorana neutrinos.
In this framework sizable light-heavy neutrino mixing is natural and sufficiently large to probe LNV at the LHC, while the active neutrinos still have sub-eV masses consistent with oscillation data. The cross section for the ($\ell^\pm \ell^\pm + 2j$) signal can be calculated from 
\begin{align}
\label{eq:lhccs-mn}
	\sigma \left(pp \to N \ell^\pm \to \ell^\pm \ell^\pm j j \right) 
	&= 
	\sigma (pp \to W \to N \ell^\pm) 
	\times \mbox{Br} (N \to \ell^\pm jj)
	\, . 
\end{align}
For a significant dilepton signal one generally requires large light-heavy mixing and heavy neutrino masses in the order of $10-100$ GeV. Within the present scenario, the branching ratio for the heavy neutrino decay is given by
\begin{align}
\label{eq:br-N-ljj}
	\mbox{Br} (N \to \ell^\pm jj) 
	=
	\frac{\Gamma(N\to \ell^\pm W)}{\Gamma^{\rm tot}_{N}}
	\times \mbox{Br} (W \to jj)
\end{align}
with $\mbox{Br} (W \to jj)=0.674$ \cite{Agashe:2014kda}. The total decay width of the heavy neutrinos $\Gamma^{\rm tot}_{N}$ is given by the sum of the following contributions
\begin{align}
	\Gamma(N\to \ell^\pm W)
	&=
	\frac{g^2_L V^2_{\ell N}}{64 \pi} \frac{M^3_N}{M^2_W} 
	\bigg(1- \frac{M^2_W}{M^2_N}\bigg)^2 \bigg(1+2 \frac{M^2_W}{M^2_N}\bigg)
	\, ,
	\label{eq:decay_2body-w}
	\\
	\Gamma(N\to \nu_\ell Z, \overline{\nu_\ell} Z)
	&=
	\frac{g^2_L V^2_{\ell N}}{128 \pi \cos^2\theta_W} \frac{M^3_N}{M^2_Z} 
	\bigg(1- \frac{M^2_Z}{M^2_N}\bigg)^2 \bigg(1+2 \frac{M^2_Z}{M^2_N}\bigg)
	\, ,
	\label{eq:decay_2body-z}
	\\
	\Gamma(N\to \nu_\ell h, \overline{\nu_\ell} h)
	&=
	\frac{g^2_L V^2_{\ell N}}{128 \pi} \frac{M^3_N}{M^2_W} 
	\bigg(1- \frac{M^2_h}{M^2_N}\bigg)^2 
	\, ,
	\label{eq:decay_2body-h}
\end{align}
where we denote by $M_N$ the mass of a heavy neutrino. The number of events expected in the dilepton channel is finally obtained from
\begin{equation}
	\#(pp \rightarrow \ell^\pm \ell^\pm jj) = \mathcal{L} \cdot \sigma \left(pp \to \ell^\pm \ell^\pm j j \right) 
	\, ,
	\label{number_of_events_in_dilepton_channel}
\end{equation}
where $\mathcal{L}$ denotes the luminosity. In \tab{\ref{tab:cross_section}} we show the results for the expected numbers of events from \eq{\ref{number_of_events_in_dilepton_channel}} for two different pairs of values of $M_N$ and $\sigma \left(pp \rightarrow N \ell^\pm \right)$ and a light-heavy mixing of $|V_{\ell N}|^2 = 10^{-4}$, where we have used the luminosity of the current and the planned LHC run \cite{CMS:2013xfa}.\footnote{Note that the values for $\sigma \left(pp \rightarrow N \ell^\pm \right)$ used here are by a factor of $10$ smaller than the ones from \cite{Chen:2011hc} corresponding to our smaller mixing. Note as well that here we have adopted the labelling of the mixing $V_{\ell N}$ of the authors of \cite{Chen:2011hc}, which is $\epsilon$ in our notation (cf$.$ \eq{\ref{mixingMatrix_orderOfMagnitude}}).} We see that in the case of a 500 GeV neutrino our model is consistent with the current measurements at the LHC, which do not observe any significant deviations from the SM. Furthermore, with about 100 events expected at a luminosity of $\mathcal{L} = 100~\text{fb}^{-1}$ the signal could be unambiguously probed in the next LHC run.
\begin{table}
\begin{center}
	\begin{tabular}{c|c|c|c}
		\hline
		\hline
		$M_N~\text{[GeV]}$	& $\sigma \left(pp \rightarrow N \ell^\pm \right)~\text{[pb]}$ & $\#$ at $\mathcal{L}= 19.4~\text{[fb]}^{-1}$   & $\#$ at $\mathcal{L}= 100~\text{[fb]}^{-1}$  \\
		\hline
		$200$	& $0.100$	& $535$ & $2760$ \\
		$500$	& $0.005$	& $23$ & $120$  \\
		\hline
		\hline
	\end{tabular}
	\caption{The expected number of events in the lepton channel for different values of $M_N$ and $\sigma \left(pp \rightarrow N \ell^\pm \right)$ \cite{Chen:2011hc} with a light-heavy mixing of $|V_{\ell N}|^2 = 10^{-4}$. The third column shows the event numbers for the current LHC run with a luminosity of $\mathcal{L}= 19.4\text{ fb}^{-1}$ and the forth the expected events in the planned LHC run with an anticipated luminosity of $\mathcal{L}= 100\text{ fb}^{-1}$.}
	\label{tab:cross_section}
\end{center}
\end{table}
\section{Conclusion \label{sec:Conclusion}}
\noindent
We have presented a novel possibility of large lepton number violation within the context of the recently explored conformal inverse seesaw mechanism. We have extended the Standard Model by additional neutrino species with a lepton number violating Majorana mass term for right-handed neutrinos as the heaviest mass scale. In the conformal framework we have introduced new scalar fields and a new gauge group that we have identified with $B - L$. The Radiative Spontaneous Symmetry Breaking has led to a hierarchy in the structure of the vacuum expectation values and the emergence of the Electro-Weak scale. We have shown that the particle spectrum of the model features active neutrinos with sub-eV scale masses in agreement with current mass limits, a keV neutrino state as a Dark Matter candidate as well as heavy neutrino states in the few GeV to hundreds of GeV range. We have demonstrated that in the model it is natural to have large mixing between the light and heavy neutrino states of the order of $10^{-2}$.

We also have discussed the phenomenological consequences of the large lepton number violation in our model in the context of neutrinoless double beta decay~($\nbb$) and the characteristic same-sign dilepton signals at the Large Hadron Collider~(LHC). We have shown that the new contributions of the heavy neutrinos to $\nbb$ can saturate the limits for the halflife in future experiment leading to a detectable signal. For the collider signatures we have estimated the expected number of events in the same-sign dilepton channel plus two jets and no missing energy. We have found that the sizable light-heavy mixing in our model can lead to a visible excess in the next LHC run. We have discussed the possibility to distinguish different mass ranges of the heavy neutrinos by analysing the kinematics of the collision products. We have commented on the same-sign dilepton channel with four jets and no missing energy mediated by the $Z^\prime$ boson associated with the $B - L$ gauge group or the mixing of the Standard Model Higgs with the new scalars as additional tests of the model at colliders. It is important to mention that, given the sizable light-heavy neutrino mixing, $\nbb$ experiments and colliders probe similar mass ranges for the heavy Majorana states. However, the couplings which are involved in the processes are different in their flavour composition and so both experiments turn out to be highly complementary. 

\section*{Acknowledgements}
The work of Sudhanwa Patra is partially supported by the Department of Science and
Technology, Govt$.$ of India under the financial grant SB/S2/HEP-011/2013 and the Max Planck Society in the project MANITOP.


\begin{thebibliography}{60}
\expandafter\ifx\csname natexlab\endcsname\relax\def\natexlab#1{#1}\fi
\expandafter\ifx\csname bibnamefont\endcsname\relax
  \def\bibnamefont#1{#1}\fi
\expandafter\ifx\csname bibfnamefont\endcsname\relax
  \def\bibfnamefont#1{#1}\fi
\expandafter\ifx\csname citenamefont\endcsname\relax
  \def\citenamefont#1{#1}\fi
\expandafter\ifx\csname url\endcsname\relax
  \def\url#1{\texttt{#1}}\fi
\expandafter\ifx\csname urlprefix\endcsname\relax\def\urlprefix{URL }\fi
\providecommand{\bibinfo}[2]{#2}
\providecommand{\eprint}[2][]{\url{#2}}

\bibitem[{\citenamefont{Kadanoff}(1971)}]{Kadanoff:1971pc}
\bibinfo{author}{\bibfnamefont{L.}~\bibnamefont{Kadanoff}},
  \bibinfo{journal}{{CRITICAL BEHAVIOR. UNIVERSALITY AND SCALING}}
  (\bibinfo{year}{1971}).

\bibitem[{\citenamefont{Coleman and Weinberg}(1973)}]{Coleman:1973jx}
\bibinfo{author}{\bibfnamefont{S.~R.} \bibnamefont{Coleman}} \bibnamefont{and}
  \bibinfo{author}{\bibfnamefont{E.~J.} \bibnamefont{Weinberg}},
  \bibinfo{journal}{Phys.Rev.} \textbf{\bibinfo{volume}{D7}},
  \bibinfo{pages}{1888} (\bibinfo{year}{1973}).


\bibitem[{\citenamefont{Hempfling}(1996)}]{Hempfling:1996ht}
\bibinfo{author}{\bibfnamefont{R.}~\bibnamefont{Hempfling}},
  \bibinfo{journal}{Phys.Lett.} \textbf{\bibinfo{volume}{B379}},
  \bibinfo{pages}{153} (\bibinfo{year}{1996}), \eprint{hep-ph/9604278}.

\bibitem[{\citenamefont{Hambye}(1996)}]{Hambye:1995fr}
\bibinfo{author}{\bibfnamefont{T.}~\bibnamefont{Hambye}},
  \bibinfo{journal}{Phys.Lett.} \textbf{\bibinfo{volume}{B371}},
  \bibinfo{pages}{87} (\bibinfo{year}{1996}), \eprint{hep-ph/9510266}.

\bibitem[{\citenamefont{Foot et~al.}(2007{\natexlab{a}})\citenamefont{Foot,
  Kobakhidze, and Volkas}}]{Foot:2007as}
\bibinfo{author}{\bibfnamefont{R.}~\bibnamefont{Foot}},
  \bibinfo{author}{\bibfnamefont{A.}~\bibnamefont{Kobakhidze}},
  \bibnamefont{and} \bibinfo{author}{\bibfnamefont{R.~R.}
  \bibnamefont{Volkas}}, \bibinfo{journal}{Phys.Lett.}
  \textbf{\bibinfo{volume}{B655}}, \bibinfo{pages}{156}
  (\bibinfo{year}{2007}{\natexlab{a}}), \eprint{0704.1165}.

\bibitem[{\citenamefont{Foot et~al.}(2007{\natexlab{b}})\citenamefont{Foot,
  Kobakhidze, McDonald, and Volkas}}]{Foot:2007ay}
\bibinfo{author}{\bibfnamefont{R.}~\bibnamefont{Foot}},
  \bibinfo{author}{\bibfnamefont{A.}~\bibnamefont{Kobakhidze}},
  \bibinfo{author}{\bibfnamefont{K.}~\bibnamefont{McDonald}}, \bibnamefont{and}
  \bibinfo{author}{\bibfnamefont{R.}~\bibnamefont{Volkas}},
  \bibinfo{journal}{Phys.Rev.} \textbf{\bibinfo{volume}{D76}},
  \bibinfo{pages}{075014} (\bibinfo{year}{2007}{\natexlab{b}}),
  \eprint{0706.1829}.

\bibitem[{\citenamefont{Chang et~al.}(2007)\citenamefont{Chang, Ng, and
  Wu}}]{Chang:2007ki}
\bibinfo{author}{\bibfnamefont{W.-F.} \bibnamefont{Chang}},
  \bibinfo{author}{\bibfnamefont{J.~N.} \bibnamefont{Ng}}, \bibnamefont{and}
  \bibinfo{author}{\bibfnamefont{J.~M.} \bibnamefont{Wu}},
  \bibinfo{journal}{Phys.Rev.} \textbf{\bibinfo{volume}{D75}},
  \bibinfo{pages}{115016} (\bibinfo{year}{2007}), \eprint{hep-ph/0701254}.

\bibitem[{\citenamefont{Hambye and Tytgat}(2008)}]{Hambye:2007vf}
\bibinfo{author}{\bibfnamefont{T.}~\bibnamefont{Hambye}} \bibnamefont{and}
  \bibinfo{author}{\bibfnamefont{M.~H.} \bibnamefont{Tytgat}},
  \bibinfo{journal}{Phys.Lett.} \textbf{\bibinfo{volume}{B659}},
  \bibinfo{pages}{651} (\bibinfo{year}{2008}), \eprint{0707.0633}.

\bibitem[{\citenamefont{Iso et~al.}(2009{\natexlab{a}})\citenamefont{Iso,
  Okada, and Orikasa}}]{Iso:2009ss}
\bibinfo{author}{\bibfnamefont{S.}~\bibnamefont{Iso}},
  \bibinfo{author}{\bibfnamefont{N.}~\bibnamefont{Okada}}, \bibnamefont{and}
  \bibinfo{author}{\bibfnamefont{Y.}~\bibnamefont{Orikasa}},
  \bibinfo{journal}{Phys.Lett.} \textbf{\bibinfo{volume}{B676}},
  \bibinfo{pages}{81} (\bibinfo{year}{2009}{\natexlab{a}}), \eprint{0902.4050}.

\bibitem[{\citenamefont{Iso et~al.}(2009{\natexlab{b}})\citenamefont{Iso,
  Okada, and Orikasa}}]{Iso:2009nw}
\bibinfo{author}{\bibfnamefont{S.}~\bibnamefont{Iso}},
  \bibinfo{author}{\bibfnamefont{N.}~\bibnamefont{Okada}}, \bibnamefont{and}
  \bibinfo{author}{\bibfnamefont{Y.}~\bibnamefont{Orikasa}},
  \bibinfo{journal}{Phys.Rev.} \textbf{\bibinfo{volume}{D80}},
  \bibinfo{pages}{115007} (\bibinfo{year}{2009}{\natexlab{b}}),
  \eprint{0909.0128}.

\bibitem[{\citenamefont{Foot et~al.}(2011)\citenamefont{Foot, Kobakhidze, and
  Volkas}}]{Foot:2010et}
\bibinfo{author}{\bibfnamefont{R.}~\bibnamefont{Foot}},
  \bibinfo{author}{\bibfnamefont{A.}~\bibnamefont{Kobakhidze}},
  \bibnamefont{and} \bibinfo{author}{\bibfnamefont{R.~R.}
  \bibnamefont{Volkas}}, \bibinfo{journal}{Phys.Rev.}
  \textbf{\bibinfo{volume}{D84}}, \bibinfo{pages}{075010}
  (\bibinfo{year}{2011}), \eprint{1012.4848}.

\bibitem[{\citenamefont{Khoze}(2013)}]{Khoze:2013uia}
\bibinfo{author}{\bibfnamefont{V.~V.} \bibnamefont{Khoze}},
  \bibinfo{journal}{JHEP} \textbf{\bibinfo{volume}{1311}}, \bibinfo{pages}{215}
  (\bibinfo{year}{2013}), \eprint{1308.6338}.

\bibitem[{\citenamefont{Kawamura}(2013)}]{Kawamura:2013kua}
\bibinfo{author}{\bibfnamefont{Y.}~\bibnamefont{Kawamura}},
  \bibinfo{journal}{PTEP} \textbf{\bibinfo{volume}{2013}},
  \bibinfo{pages}{113B04} (\bibinfo{year}{2013}), \eprint{1308.5069}.

\bibitem[{\citenamefont{Gretsch and Monin}(2013)}]{Gretsch:2013ooa}
\bibinfo{author}{\bibfnamefont{F.}~\bibnamefont{Gretsch}} \bibnamefont{and}
  \bibinfo{author}{\bibfnamefont{A.}~\bibnamefont{Monin}}
  (\bibinfo{year}{2013}), \eprint{1308.3863}.

\bibitem[{\citenamefont{Heikinheimo et~al.}(2014)\citenamefont{Heikinheimo,
  Racioppi, Raidal, Spethmann, and Tuominen}}]{Heikinheimo:2013fta}
\bibinfo{author}{\bibfnamefont{M.}~\bibnamefont{Heikinheimo}},
  \bibinfo{author}{\bibfnamefont{A.}~\bibnamefont{Racioppi}},
  \bibinfo{author}{\bibfnamefont{M.}~\bibnamefont{Raidal}},
  \bibinfo{author}{\bibfnamefont{C.}~\bibnamefont{Spethmann}},
  \bibnamefont{and} \bibinfo{author}{\bibfnamefont{K.}~\bibnamefont{Tuominen}},
  \bibinfo{journal}{Mod.Phys.Lett.} \textbf{\bibinfo{volume}{A29}},
  \bibinfo{pages}{1450077} (\bibinfo{year}{2014}), \eprint{1304.7006}.

\bibitem[{\citenamefont{Gabrielli et~al.}(2014)\citenamefont{Gabrielli,
  Heikinheimo, Kannike, Racioppi, Raidal et~al.}}]{Gabrielli:2013hma}
\bibinfo{author}{\bibfnamefont{E.}~\bibnamefont{Gabrielli}},
  \bibinfo{author}{\bibfnamefont{M.}~\bibnamefont{Heikinheimo}},
  \bibinfo{author}{\bibfnamefont{K.}~\bibnamefont{Kannike}},
  \bibinfo{author}{\bibfnamefont{A.}~\bibnamefont{Racioppi}},
  \bibinfo{author}{\bibfnamefont{M.}~\bibnamefont{Raidal}},
  \bibnamefont{et~al.}, \bibinfo{journal}{Phys.Rev.}
  \textbf{\bibinfo{volume}{D89}}, \bibinfo{pages}{015017}
  (\bibinfo{year}{2014}), \eprint{1309.6632}.

\bibitem[{\citenamefont{Carone and Ramos}(2013)}]{Carone:2013wla}
\bibinfo{author}{\bibfnamefont{C.~D.} \bibnamefont{Carone}} \bibnamefont{and}
  \bibinfo{author}{\bibfnamefont{R.}~\bibnamefont{Ramos}},
  \bibinfo{journal}{Phys.Rev.} \textbf{\bibinfo{volume}{D88}},
  \bibinfo{pages}{055020} (\bibinfo{year}{2013}), \eprint{1307.8428}.

\bibitem[{\citenamefont{Khoze and Ro}(2013)}]{Khoze:2013oga}
\bibinfo{author}{\bibfnamefont{V.~V.} \bibnamefont{Khoze}} \bibnamefont{and}
  \bibinfo{author}{\bibfnamefont{G.}~\bibnamefont{Ro}}, \bibinfo{journal}{JHEP}
  \textbf{\bibinfo{volume}{1310}}, \bibinfo{pages}{075} (\bibinfo{year}{2013}),
  \eprint{1307.3764}.

\bibitem[{\citenamefont{Englert et~al.}(2013)\citenamefont{Englert, Jaeckel,
  Khoze, and Spannowsky}}]{Englert:2013gz}
\bibinfo{author}{\bibfnamefont{C.}~\bibnamefont{Englert}},
  \bibinfo{author}{\bibfnamefont{J.}~\bibnamefont{Jaeckel}},
  \bibinfo{author}{\bibfnamefont{V.}~\bibnamefont{Khoze}}, \bibnamefont{and}
  \bibinfo{author}{\bibfnamefont{M.}~\bibnamefont{Spannowsky}},
  \bibinfo{journal}{JHEP} \textbf{\bibinfo{volume}{1304}}, \bibinfo{pages}{060}
  (\bibinfo{year}{2013}), \eprint{1301.4224}.

\bibitem[{\citenamefont{Farzinnia et~al.}(2013)\citenamefont{Farzinnia, He, and
  Ren}}]{Farzinnia:2013pga}
\bibinfo{author}{\bibfnamefont{A.}~\bibnamefont{Farzinnia}},
  \bibinfo{author}{\bibfnamefont{H.-J.} \bibnamefont{He}}, \bibnamefont{and}
  \bibinfo{author}{\bibfnamefont{J.}~\bibnamefont{Ren}},
  \bibinfo{journal}{Phys.Lett.} \textbf{\bibinfo{volume}{B727}},
  \bibinfo{pages}{141} (\bibinfo{year}{2013}), \eprint{1308.0295}.

\bibitem[{\citenamefont{Abel and Mariotti}(2013)}]{Abel:2013mya}
\bibinfo{author}{\bibfnamefont{S.}~\bibnamefont{Abel}} \bibnamefont{and}
  \bibinfo{author}{\bibfnamefont{A.}~\bibnamefont{Mariotti}}
  (\bibinfo{year}{2013}), \eprint{1312.5335}.

\bibitem[{\citenamefont{Foot et~al.}(2013)\citenamefont{Foot, Kobakhidze,
  McDonald, and Volkas}}]{Foot:2013hna}
\bibinfo{author}{\bibfnamefont{R.}~\bibnamefont{Foot}},
  \bibinfo{author}{\bibfnamefont{A.}~\bibnamefont{Kobakhidze}},
  \bibinfo{author}{\bibfnamefont{K.~L.} \bibnamefont{McDonald}},
  \bibnamefont{and} \bibinfo{author}{\bibfnamefont{R.~R.} \bibnamefont{Volkas}}
  (\bibinfo{year}{2013}), \eprint{1310.0223}.

\bibitem[{\citenamefont{Hill}(2014)}]{Hill:2014mqa}
\bibinfo{author}{\bibfnamefont{C.~T.} \bibnamefont{Hill}},
  \bibinfo{journal}{Phys.Rev.} \textbf{\bibinfo{volume}{D89}},
  \bibinfo{pages}{073003} (\bibinfo{year}{2014}), \eprint{1401.4185}.

\bibitem[{\citenamefont{Guo and Kang}(2014)}]{Guo:2014bha}
\bibinfo{author}{\bibfnamefont{J.}~\bibnamefont{Guo}} \bibnamefont{and}
  \bibinfo{author}{\bibfnamefont{Z.}~\bibnamefont{Kang}}
  (\bibinfo{year}{2014}), \eprint{1401.5609}.

\bibitem[{\citenamefont{Alexander-Nunneley and
  Pilaftsis}(2010)}]{AlexanderNunneley:2010nw}
\bibinfo{author}{\bibfnamefont{L.}~\bibnamefont{Alexander-Nunneley}}
  \bibnamefont{and}
  \bibinfo{author}{\bibfnamefont{A.}~\bibnamefont{Pilaftsis}},
  \bibinfo{journal}{JHEP} \textbf{\bibinfo{volume}{1009}}, \bibinfo{pages}{021}
  (\bibinfo{year}{2010}), \eprint{1006.5916}.

\bibitem[{\citenamefont{Radovcic and Benic}(2014)}]{Radovcic:2014rea}
\bibinfo{author}{\bibfnamefont{B.}~\bibnamefont{Radovcic}} \bibnamefont{and}
  \bibinfo{author}{\bibfnamefont{S.}~\bibnamefont{Benic}}
  (\bibinfo{year}{2014}), \eprint{1401.8183}.

\bibitem[{\citenamefont{Khoze et~al.}(2014)\citenamefont{Khoze, McCabe, and
  Ro}}]{Khoze:2014xha}
\bibinfo{author}{\bibfnamefont{V.~V.} \bibnamefont{Khoze}},
  \bibinfo{author}{\bibfnamefont{C.}~\bibnamefont{McCabe}}, \bibnamefont{and}
  \bibinfo{author}{\bibfnamefont{G.}~\bibnamefont{Ro}} (\bibinfo{year}{2014}),
  \eprint{1403.4953}.

\bibitem[{\citenamefont{Smirnov}(2014)}]{Smirnov:2014zga}
\bibinfo{author}{\bibfnamefont{J.}~\bibnamefont{Smirnov}}
  (\bibinfo{year}{2014}), \eprint{1402.1490}.

\bibitem[{\citenamefont{Salvio and Strumia}(2014)}]{Salvio:2014soa}
\bibinfo{author}{\bibfnamefont{A.}~\bibnamefont{Salvio}} \bibnamefont{and}
  \bibinfo{author}{\bibfnamefont{A.}~\bibnamefont{Strumia}},
  \bibinfo{journal}{JHEP} \textbf{\bibinfo{volume}{1406}}, \bibinfo{pages}{080}
  (\bibinfo{year}{2014}), \eprint{1403.4226}.

\bibitem[{\citenamefont{Chankowski et~al.}(2014)\citenamefont{Chankowski,
  Lewandowski, Meissner, and Nicolai}}]{Chankowski:2014fva}
\bibinfo{author}{\bibfnamefont{P.~H.} \bibnamefont{Chankowski}},
  \bibinfo{author}{\bibfnamefont{A.}~\bibnamefont{Lewandowski}},
  \bibinfo{author}{\bibfnamefont{K.~A.} \bibnamefont{Meissner}},
  \bibnamefont{and} \bibinfo{author}{\bibfnamefont{H.}~\bibnamefont{Nicolai}}
  (\bibinfo{year}{2014}), \eprint{1404.0548}.

\bibitem[{\citenamefont{Okada and Orikasa}(2014{\natexlab{a}})}]{Okada:2014nea}
\bibinfo{author}{\bibfnamefont{H.}~\bibnamefont{Okada}} \bibnamefont{and}
  \bibinfo{author}{\bibfnamefont{Y.}~\bibnamefont{Orikasa}}
  (\bibinfo{year}{2014}{\natexlab{a}}), \eprint{1412.3616}.

\bibitem[{\citenamefont{Guo et~al.}(2015)\citenamefont{Guo, Kang, Ko, and
  Orikasa}}]{Guo:2015lxa}
\bibinfo{author}{\bibfnamefont{J.}~\bibnamefont{Guo}},
  \bibinfo{author}{\bibfnamefont{Z.}~\bibnamefont{Kang}},
  \bibinfo{author}{\bibfnamefont{P.}~\bibnamefont{Ko}}, \bibnamefont{and}
  \bibinfo{author}{\bibfnamefont{Y.}~\bibnamefont{Orikasa}}
  (\bibinfo{year}{2015}), \eprint{1502.00508}.

\bibitem[{\citenamefont{Baek et~al.}(2015)\citenamefont{Baek, Okada, and
  Yagyu}}]{Baek:2015mna}
\bibinfo{author}{\bibfnamefont{S.}~\bibnamefont{Baek}},
  \bibinfo{author}{\bibfnamefont{H.}~\bibnamefont{Okada}}, \bibnamefont{and}
  \bibinfo{author}{\bibfnamefont{K.}~\bibnamefont{Yagyu}}
  (\bibinfo{year}{2015}), \eprint{1501.01530}.

\bibitem[{\citenamefont{Hatanaka et~al.}(2014)\citenamefont{Hatanaka,
  Nishiwaki, Okada, and Orikasa}}]{Hatanaka:2014tba}
\bibinfo{author}{\bibfnamefont{H.}~\bibnamefont{Hatanaka}},
  \bibinfo{author}{\bibfnamefont{K.}~\bibnamefont{Nishiwaki}},
  \bibinfo{author}{\bibfnamefont{H.}~\bibnamefont{Okada}}, \bibnamefont{and}
  \bibinfo{author}{\bibfnamefont{Y.}~\bibnamefont{Orikasa}}
  (\bibinfo{year}{2014}), \eprint{1412.8664}.

\bibitem[{\citenamefont{Benic and Radovcic}(2015)}]{Benic:2014aga}
\bibinfo{author}{\bibfnamefont{S.}~\bibnamefont{Benic}} \bibnamefont{and}
  \bibinfo{author}{\bibfnamefont{B.}~\bibnamefont{Radovcic}},
  \bibinfo{journal}{JHEP} \textbf{\bibinfo{volume}{1501}}, \bibinfo{pages}{143}
  (\bibinfo{year}{2015}), \eprint{1409.5776}.

\bibitem[{\citenamefont{Gorsky et~al.}(2014)\citenamefont{Gorsky, Mironov,
  Morozov, and Tomaras}}]{Gorsky:2014una}
\bibinfo{author}{\bibfnamefont{A.}~\bibnamefont{Gorsky}},
  \bibinfo{author}{\bibfnamefont{A.}~\bibnamefont{Mironov}},
  \bibinfo{author}{\bibfnamefont{A.}~\bibnamefont{Morozov}}, \bibnamefont{and}
  \bibinfo{author}{\bibfnamefont{T.}~\bibnamefont{Tomaras}}
  (\bibinfo{year}{2014}), \eprint{1409.0492}.

\bibitem[{\citenamefont{Okada et~al.}(2014)\citenamefont{Okada, Toma, and
  Yagyu}}]{Okada:2014qsa}
\bibinfo{author}{\bibfnamefont{H.}~\bibnamefont{Okada}},
  \bibinfo{author}{\bibfnamefont{T.}~\bibnamefont{Toma}}, \bibnamefont{and}
  \bibinfo{author}{\bibfnamefont{K.}~\bibnamefont{Yagyu}},
  \bibinfo{journal}{Phys.Rev.} \textbf{\bibinfo{volume}{D90}},
  \bibinfo{pages}{095005} (\bibinfo{year}{2014}), \eprint{1408.0961}.

\bibitem[{\citenamefont{Okada and Orikasa}(2014{\natexlab{b}})}]{Okada:2014oda}
\bibinfo{author}{\bibfnamefont{H.}~\bibnamefont{Okada}} \bibnamefont{and}
  \bibinfo{author}{\bibfnamefont{Y.}~\bibnamefont{Orikasa}},
  \bibinfo{journal}{Phys.Rev.} \textbf{\bibinfo{volume}{D90}},
  \bibinfo{pages}{075023} (\bibinfo{year}{2014}{\natexlab{b}}),
  \eprint{1407.2543}.

\bibitem[{\citenamefont{Khoze and Ro}(2014)}]{Khoze:2014woa}
\bibinfo{author}{\bibfnamefont{V.~V.} \bibnamefont{Khoze}} \bibnamefont{and}
  \bibinfo{author}{\bibfnamefont{G.}~\bibnamefont{Ro}}, \bibinfo{journal}{JHEP}
  \textbf{\bibinfo{volume}{1410}}, \bibinfo{pages}{61} (\bibinfo{year}{2014}),
  \eprint{1406.2291}.

\bibitem[{\citenamefont{Lattanzi et~al.}(2014)\citenamefont{Lattanzi, Lineros,
  and Taoso}}]{Lattanzi:2014mia}
\bibinfo{author}{\bibfnamefont{M.}~\bibnamefont{Lattanzi}},
  \bibinfo{author}{\bibfnamefont{R.~A.} \bibnamefont{Lineros}},
  \bibnamefont{and} \bibinfo{author}{\bibfnamefont{M.}~\bibnamefont{Taoso}},
  \bibinfo{journal}{New J.Phys.} \textbf{\bibinfo{volume}{16}},
  \bibinfo{pages}{125012} (\bibinfo{year}{2014}), \eprint{1406.0004}.

\bibitem[{\citenamefont{Kallosh et~al.}(1995)\citenamefont{Kallosh, Linde,
  Linde, and Susskind}}]{Kallosh:1995hi}
\bibinfo{author}{\bibfnamefont{R.}~\bibnamefont{Kallosh}},
  \bibinfo{author}{\bibfnamefont{A.~D.} \bibnamefont{Linde}},
  \bibinfo{author}{\bibfnamefont{D.~A.} \bibnamefont{Linde}}, \bibnamefont{and}
  \bibinfo{author}{\bibfnamefont{L.}~\bibnamefont{Susskind}},
  \bibinfo{journal}{Phys.Rev.} \textbf{\bibinfo{volume}{D52}},
  \bibinfo{pages}{912} (\bibinfo{year}{1995}), \eprint{hep-th/9502069}.

\bibitem[{\citenamefont{Humbert et~al.}(2015)\citenamefont{Humbert, Lindner,
  and Smirnov}}]{Humbert:2015epa}
\bibinfo{author}{\bibfnamefont{P.}~\bibnamefont{Humbert}},
  \bibinfo{author}{\bibfnamefont{M.}~\bibnamefont{Lindner}}, \bibnamefont{and}
  \bibinfo{author}{\bibfnamefont{J.}~\bibnamefont{Smirnov}}
  (\bibinfo{year}{2015}), \eprint{1503.03066}.

\bibitem[{\citenamefont{Minkowski}(1977)}]{Minkowski:1977sc}
\bibinfo{author}{\bibfnamefont{P.}~\bibnamefont{Minkowski}},
  \bibinfo{journal}{Phys.Lett.} \textbf{\bibinfo{volume}{B67}},
  \bibinfo{pages}{421} (\bibinfo{year}{1977}).

\bibitem[{\citenamefont{Mohapatra and Senjanovic}(1980)}]{Mohapatra:1979ia}
\bibinfo{author}{\bibfnamefont{R.~N.} \bibnamefont{Mohapatra}}
  \bibnamefont{and}
  \bibinfo{author}{\bibfnamefont{G.}~\bibnamefont{Senjanovic}},
  \bibinfo{journal}{Phys.Rev.Lett.} \textbf{\bibinfo{volume}{44}},
  \bibinfo{pages}{912} (\bibinfo{year}{1980}).

\bibitem[{\citenamefont{Mohapatra and Valle}(1986)}]{Mohapatra:1986bd}
\bibinfo{author}{\bibfnamefont{R.}~\bibnamefont{Mohapatra}} \bibnamefont{and}
  \bibinfo{author}{\bibfnamefont{J.}~\bibnamefont{Valle}},
  \bibinfo{journal}{Phys.Rev.} \textbf{\bibinfo{volume}{D34}},
  \bibinfo{pages}{1642} (\bibinfo{year}{1986}).

\bibitem[{\citenamefont{Deppisch and Valle}(2005)}]{Deppisch:2004fa}
\bibinfo{author}{\bibfnamefont{F.}~\bibnamefont{Deppisch}} \bibnamefont{and}
  \bibinfo{author}{\bibfnamefont{J.}~\bibnamefont{Valle}},
  \bibinfo{journal}{Phys.Rev.} \textbf{\bibinfo{volume}{D72}},
  \bibinfo{pages}{036001} (\bibinfo{year}{2005}), \eprint{hep-ph/0406040}.

\bibitem[{\citenamefont{Lindner et~al.}(2014)\citenamefont{Lindner, Schmidt,
  and Smirnov}}]{Lindner:2014oea}
\bibinfo{author}{\bibfnamefont{M.}~\bibnamefont{Lindner}},
  \bibinfo{author}{\bibfnamefont{S.}~\bibnamefont{Schmidt}}, \bibnamefont{and}
  \bibinfo{author}{\bibfnamefont{J.}~\bibnamefont{Smirnov}},
  \bibinfo{journal}{JHEP} \textbf{\bibinfo{volume}{1410}}, \bibinfo{pages}{177}
  (\bibinfo{year}{2014}), \eprint{1405.6204}.

\bibitem[{\citenamefont{Bulbul et~al.}(2014)\citenamefont{Bulbul, Markevitch,
  Foster, Smith, Loewenstein et~al.}}]{Bulbul:2014sua}
\bibinfo{author}{\bibfnamefont{E.}~\bibnamefont{Bulbul}},
  \bibinfo{author}{\bibfnamefont{M.}~\bibnamefont{Markevitch}},
  \bibinfo{author}{\bibfnamefont{A.}~\bibnamefont{Foster}},
  \bibinfo{author}{\bibfnamefont{R.~K.} \bibnamefont{Smith}},
  \bibinfo{author}{\bibfnamefont{M.}~\bibnamefont{Loewenstein}},
  \bibnamefont{et~al.}, \bibinfo{journal}{Astrophys.J.}
  \textbf{\bibinfo{volume}{789}}, \bibinfo{pages}{13} (\bibinfo{year}{2014}),
  \eprint{1402.2301}.

\bibitem[{\citenamefont{Merle and Totzauer}(2015)}]{Merle:2015oja}
\bibinfo{author}{\bibfnamefont{A.}~\bibnamefont{Merle}} \bibnamefont{and}
  \bibinfo{author}{\bibfnamefont{M.}~\bibnamefont{Totzauer}}
  (\bibinfo{year}{2015}), \eprint{1502.01011}.

\bibitem[{\citenamefont{Meroni et~al.}(2013)\citenamefont{Meroni, Petcov, and
  Simkovic}}]{Meroni:2012qf}
\bibinfo{author}{\bibfnamefont{A.}~\bibnamefont{Meroni}},
  \bibinfo{author}{\bibfnamefont{S.}~\bibnamefont{Petcov}}, \bibnamefont{and}
  \bibinfo{author}{\bibfnamefont{F.}~\bibnamefont{Simkovic}},
  \bibinfo{journal}{JHEP} \textbf{\bibinfo{volume}{1302}}, \bibinfo{pages}{025}
  (\bibinfo{year}{2013}), \eprint{1212.1331}.

\bibitem[{\citenamefont{Agostini et~al.}(2013)}]{Agostini:2013mzu}
\bibinfo{author}{\bibfnamefont{M.}~\bibnamefont{Agostini}} \bibnamefont{et~al.}
  (\bibinfo{collaboration}{GERDA}), \bibinfo{journal}{Phys.Rev.Lett.}
  \textbf{\bibinfo{volume}{111}}, \bibinfo{pages}{122503}
  (\bibinfo{year}{2013}), \eprint{1307.4720}.

\bibitem[{\citenamefont{Auger et~al.}(2012)}]{Auger:2012ar}
\bibinfo{author}{\bibfnamefont{M.}~\bibnamefont{Auger}} \bibnamefont{et~al.}
  (\bibinfo{collaboration}{EXO}), \bibinfo{journal}{Phys.Rev.Lett.}
  \textbf{\bibinfo{volume}{109}}, \bibinfo{pages}{032505}
  (\bibinfo{year}{2012}), \eprint{1205.5608}.

\bibitem[{\citenamefont{Gando et~al.}(2013)}]{Gando:2012zm}
\bibinfo{author}{\bibfnamefont{A.}~\bibnamefont{Gando}} \bibnamefont{et~al.}
  (\bibinfo{collaboration}{KamLAND-Zen}), \bibinfo{journal}{Phys.Rev.Lett.}
  \textbf{\bibinfo{volume}{110}}, \bibinfo{pages}{062502}
  (\bibinfo{year}{2013}), \eprint{1211.3863}.

\bibitem[{\citenamefont{Izaguirre and Shuve}(2015)}]{Izaguirre:2015pga}
\bibinfo{author}{\bibfnamefont{E.}~\bibnamefont{Izaguirre}} \bibnamefont{and}
  \bibinfo{author}{\bibfnamefont{B.}~\bibnamefont{Shuve}}
  (\bibinfo{year}{2015}), \eprint{1504.02470}.

\bibitem[{\citenamefont{Almeida et~al.}(2000)\citenamefont{Almeida, Coutinho,
  Martins~Simoes, and do~Vale}}]{Almeida:2000pz}
\bibinfo{author}{\bibfnamefont{J.}~\bibnamefont{Almeida},
  \bibfnamefont{F.M.L.}}, \bibinfo{author}{\bibfnamefont{Y.~D.~A.}
  \bibnamefont{Coutinho}}, \bibinfo{author}{\bibfnamefont{J.~A.}
  \bibnamefont{Martins~Simoes}}, \bibnamefont{and}
  \bibinfo{author}{\bibfnamefont{M.}~\bibnamefont{do~Vale}},
  \bibinfo{journal}{Phys.Rev.} \textbf{\bibinfo{volume}{D62}},
  \bibinfo{pages}{075004} (\bibinfo{year}{2000}), \eprint{hep-ph/0002024}.

\bibitem[{\citenamefont{Olive et~al.}(2014)}]{Agashe:2014kda}
\bibinfo{author}{\bibfnamefont{K.}~\bibnamefont{Olive}} \bibnamefont{et~al.}
  (\bibinfo{collaboration}{Particle Data Group}), \bibinfo{journal}{Chin.Phys.}
  \textbf{\bibinfo{volume}{C38}}, \bibinfo{pages}{090001}
  (\bibinfo{year}{2014}).

\bibitem[{\citenamefont{1244669}(2013)}]{CMS:2013xfa}
\bibinfo{author}{\bibnamefont{1244669}} (\bibinfo{collaboration}{CMS
  Collaboration}) (\bibinfo{year}{2013}), \eprint{1307.7135}.

\bibitem[{\citenamefont{Chen and Dev}(2012)}]{Chen:2011hc}
\bibinfo{author}{\bibfnamefont{C.-Y.} \bibnamefont{Chen}} \bibnamefont{and}
  \bibinfo{author}{\bibfnamefont{P.~B.} \bibnamefont{Dev}},
  \bibinfo{journal}{Phys.Rev.} \textbf{\bibinfo{volume}{D85}},
  \bibinfo{pages}{093018} (\bibinfo{year}{2012}), \eprint{1112.6419}.

\end{thebibliography}
\end{document}